\documentclass{PoS}

\title{Study of silicon photomultipliers for the readout of scintillator crystals in the proposed GRIPS $\gamma$--ray astronomy mission}

\ShortTitle{SiPMs for GRIPS}
\author{\speaker{Alexei Ulyanov}, Lorraine Hanlon, Sheila McBreen, Suzanne Foley and David Byrne \\
Space Science Group, School of Physics, University College Dublin, Ireland\\
E-mail: \email{alexey.uliyanov@ucd.ie }
}

\abstract{Among the top priorities for high-energy astronomy in the coming decade are sensitive surveys in the hard X-ray/soft $\gamma$--ray (10--600 keV) and medium-energy $\gamma$--ray (0.2--80 MeV) bands. Historically, observations in the soft and medium energy $\gamma$--ray bands have been conducted using detectors based on inorganic scintillators read out by photo-multiplier tubes (PMTs). These observations were limited by the modest energy and time resolution of traditional scintillator materials (e.g. NaI and CsI), and by the demands on mission resources imposed by the bulky, fragile, high-voltage PMTs. Recent technological advances in the development of both new scintillator materials (e.g. LaBr$_3$:Ce, L(Y)SO) and new scintillation light readout devices (e.g. Silicon Photo-Multipliers) promise to greatly improve the observational capabilities of future scintillator-based $\gamma$--ray telescopes, while retaining the relative simplicity, reliability, large collection volumes, and low-cost of scintillator instruments. We present initial results of a study on the use of silicon photomultipliers in the calorimeter module design of the proposed GRIPS astrophysics mission. }
\FullConference{``An INTEGRAL view of the high-energy sky (the first 10 years)"
9th INTEGRAL Workshop and celebration of the 10th anniversary of the launch,\\
		October 15-19, 2012\\
		Bibliotheque Nationale de France, Paris, France }

\usepackage{graphicx}
\usepackage[absolute]{textpos}
\usepackage{multicol}

\usepackage{color}
\definecolor{Orchid}{cmyk}{0.32,0.64,0,0}
\definecolor{Dark}{gray}{.20}
\definecolor{cafecomleite}{cmyk}{0.00,0.03,0.09,0.00}

\begin{document}


\maketitle
\vspace*{-.6cm}
\thispagestyle{empty}
\section{Introduction}

GRIPS (Gamma-Ray Imaging, Polarimetry and Spectroscopy)~\cite{greiner2009, greiner} is a proposed $\gamma$-ray astronomy mission, which will perform a sensitive all-sky scanning survey from 200 keV to 80 MeV. With respect to previous missions (e.g. INTEGRAL, COMPTEL, EGRET) the sensitivity in this energy range will be improved by at least an order of magnitude. The GRIPS mission will investigate $\gamma$-ray bursts and blazars, the mechanisms behind supernova explosions, nucleosynthesis and spallation, the origin of positrons in our Galaxy, and the nature of radiation processes and particle acceleration in extreme cosmic sources including pulsars and magnetars.

As its primary instrument, GRIPS will carry a combined Compton scattering and pair creation telescope called the Gamma-Ray Monitor (GRM).
Similar to previous Compton and pair creation telescopes, the GRM design envisages two separate detectors: a silicon tracker, in which the initial Compton scattering or pair conversion takes place, and a calorimeter, which absorbs and measures the energy of the secondaries. The calorimeter employs LaBr$_3$ scintillator crystals to achieve an improved energy resolution (about 3\% at 1~MeV) and faster response times compared to traditional scintillator materials such as NaI, CsI, or BGO.

For optimum angular resolution, GRM requires a finely segmented calorimeter with $\sim 10^5$ readout channels. This precludes the use of bulky, fragile, high-voltage photomultiplier tubes, traditionally employed for the readout of scintillator crystals. PIN diodes are not suitable because they have no internal gain and produce signals with a low signal-to-noise ratio, resulting in poor energy resolution. The current GRM baseline design relies on the development of advanced semiconductor photodetectors such as silicon drift detectors (SDD). Recent technological advances in the development of silicon photomultipliers (SiPM) make them another promising alternative for the GRM calorimeter readout ~\cite{garutti2011}.  These detectors combine the high gain of traditional photomultipliers with low-voltage operation, robustness, low mass and compact design typical for semiconductor devices. The purpose of the present study is to evaluate the performance of an SiPM-based calorimeter readout in the GRM and to optimise the design of a calorimeter module in order to meet the baseline performance requirements. This work is a collaboration between the Space Science Group in University College Dublin and SensL Technologies Ltd (http://www.sensl.com), an Irish company who have developed and will supply the SiPMs. This paper focusses on the simulation of the optical performance of the SiPM and the transport of light within a calorimeter module.


\section{Silicon photomultipliers}

\begin{figure}
\resizebox{!}{4.5cm}{\includegraphics{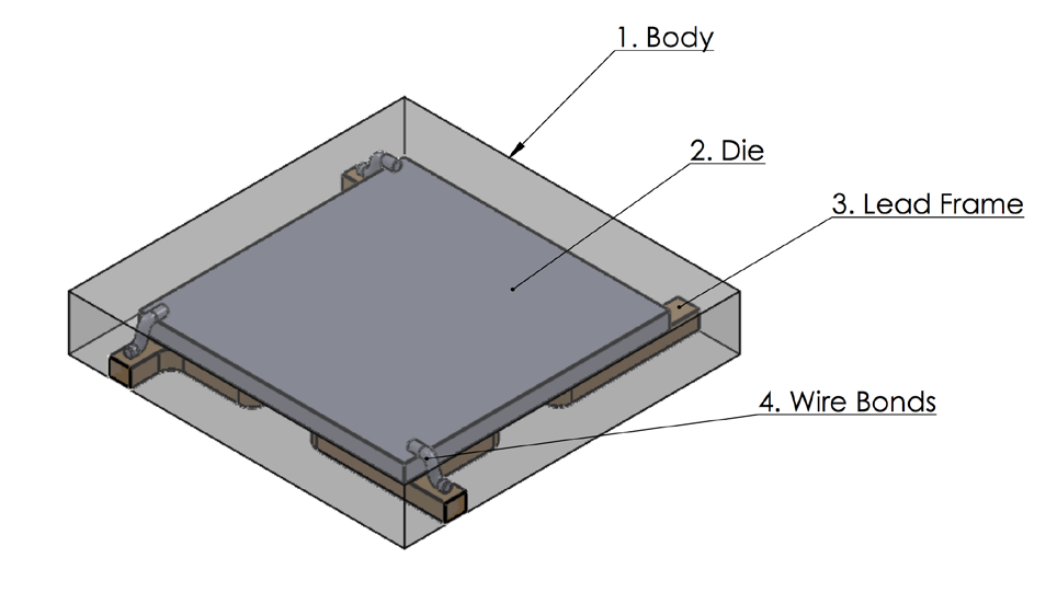}}
\resizebox{!}{4.5cm}{\includegraphics{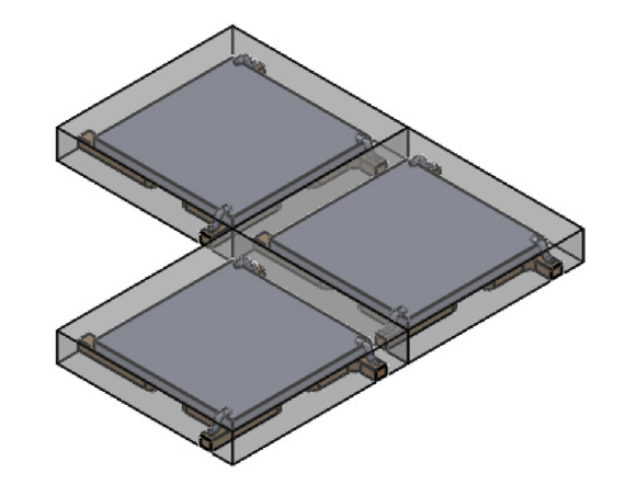}}
\caption{Silicon photomultiplier SMT package from SensL (\textbf{left}) and tiling example (\textbf{right}). Drawings by SensL.}
\label{fig1}
\end{figure}

A new generation of SensL SiPMs, optimised for blue light sensitivity, is close to market readiness. The company has developed a custom surface-mount
technology (SMT) package for the new devices that is compact, leadless and is compatible with lead-free, reflow soldering processes (Figure~\ref{fig1}). A clear
encapsulant is used to provide a smooth flat surface for optimal coupling to scintillators.
The dead space between the detector chip and the edge of the package has been minimised resulting in a package that can be tiled
on four sides. This allows an array of multiple devices to be assembled
on a PCB (Figure~\ref{fig1}). The distance between the active areas of tiled devices is about 1~mm.

%


An SiPM essentially consists of a large number of avalanche photodiodes (microcells) operating in Geiger mode. When $N_{\rm ph}$ photons hit the surface of a silicon photomultiplier, the average number of photons that trigger avalanches (primary triggers)
is defined by the photon detection efficiency (PDE):
\[
<N_{\rm trig}>={\rm PDE}\cdot N_{\rm ph}
\]
Additional avalanche triggers are generated by thermally produced electrons and holes (dark noise),
photons produced in the avalanches and propagating to the adjacent microcells
(cross-talk)  and trapped charge carriers (after-pulses).

\begin{figure}[ht]
\centering
    \includegraphics[totalheight=0.2\textheight]{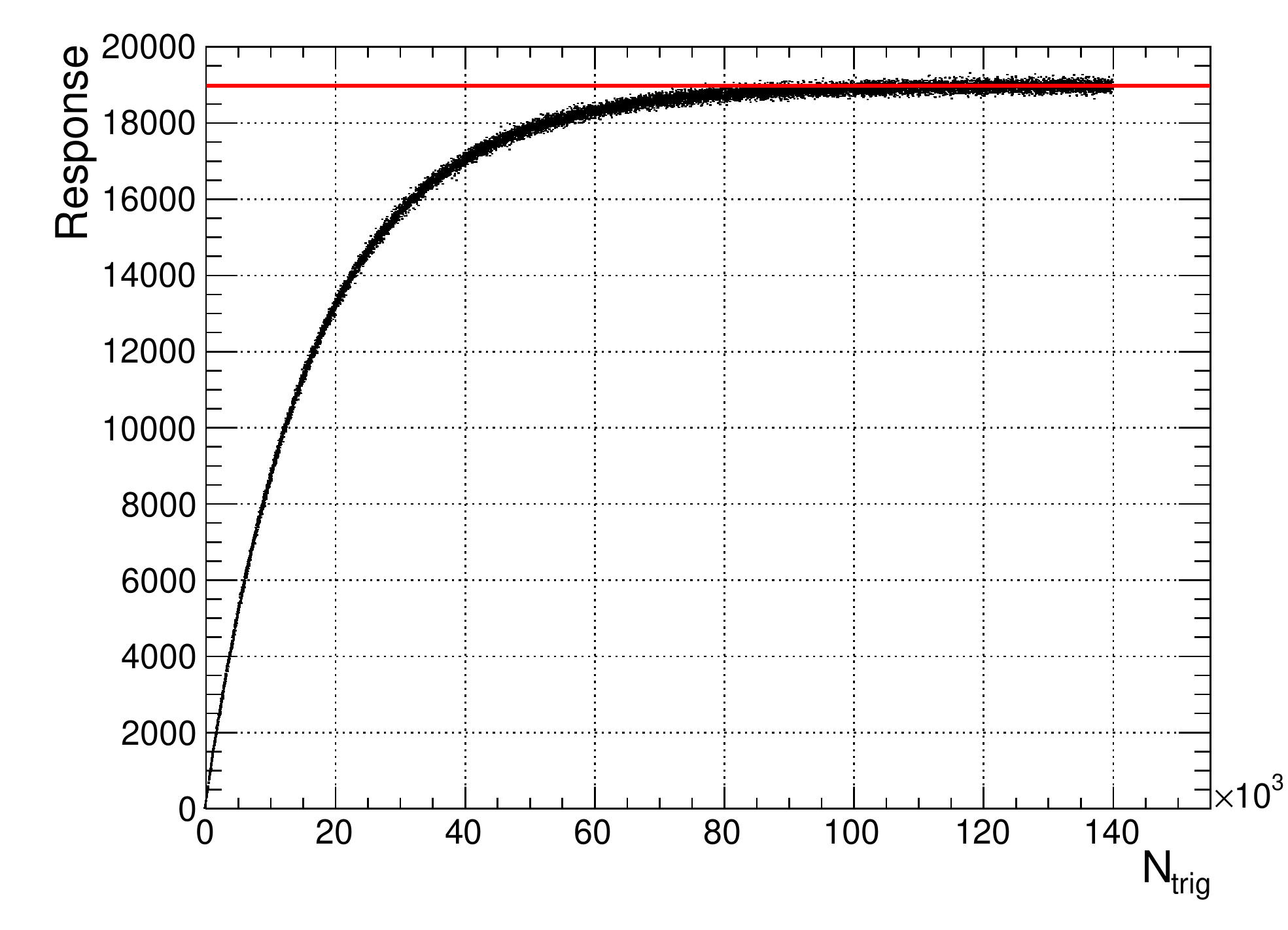}
    \includegraphics[totalheight=0.2\textheight]{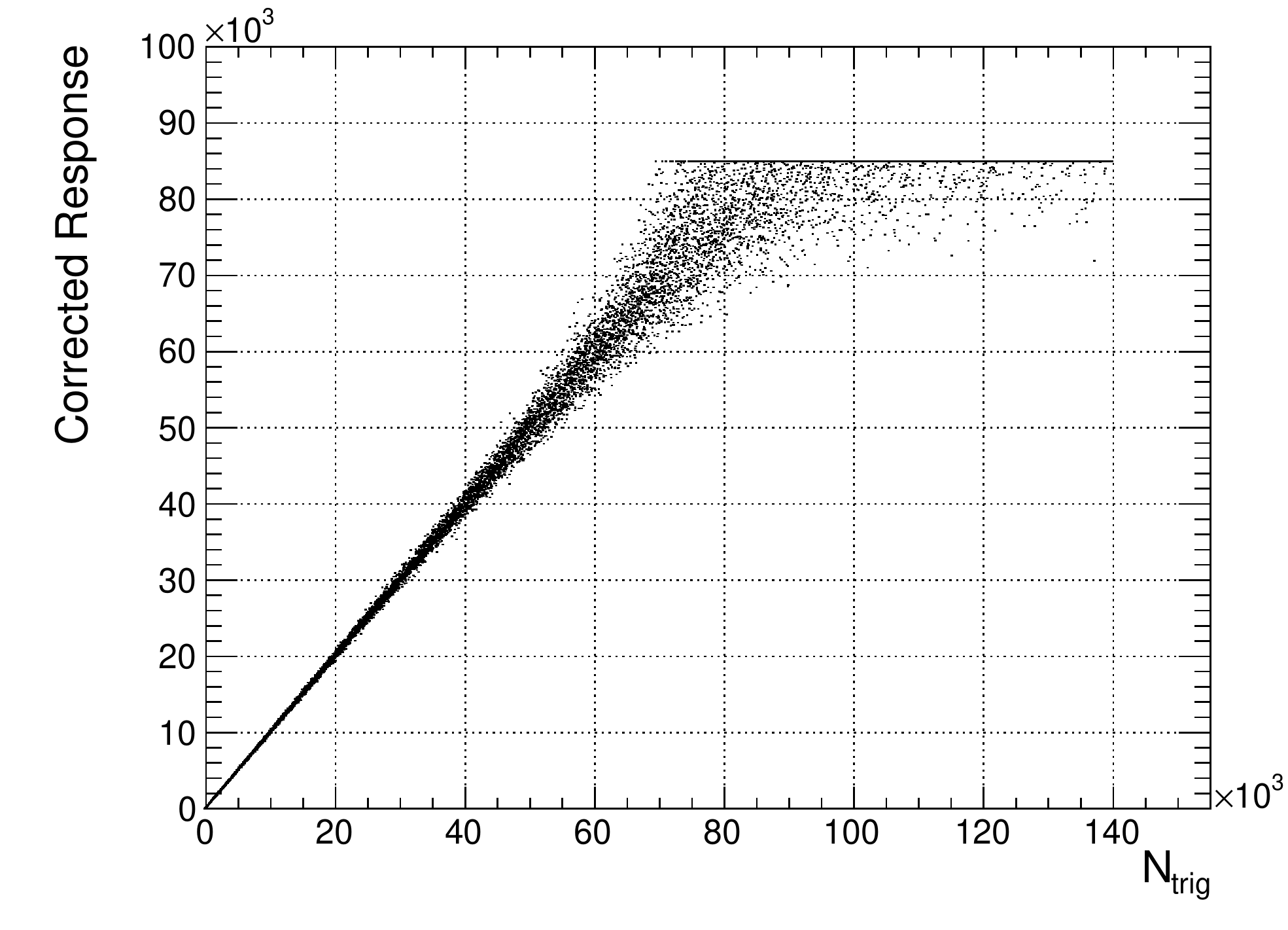}
    \caption{SiPM simulated response to light versus the number of triggering photons. \textbf{Left:} The raw detector response divided by the average charge released by a single microcell. The response is limited by the total number of
microcells (red line). Visible response fluctuations are mostly due to the
assumed 0.5\% gain fluctuations. \textbf{Right:} Response corrections are used to linearise the detector response. When the raw detector response approaches
its limit, the corrected response is fixed at the maximum value.      }
\label{fig:response}
\end{figure}

\begin{figure}
\centering
    \includegraphics[totalheight=0.2\textheight]{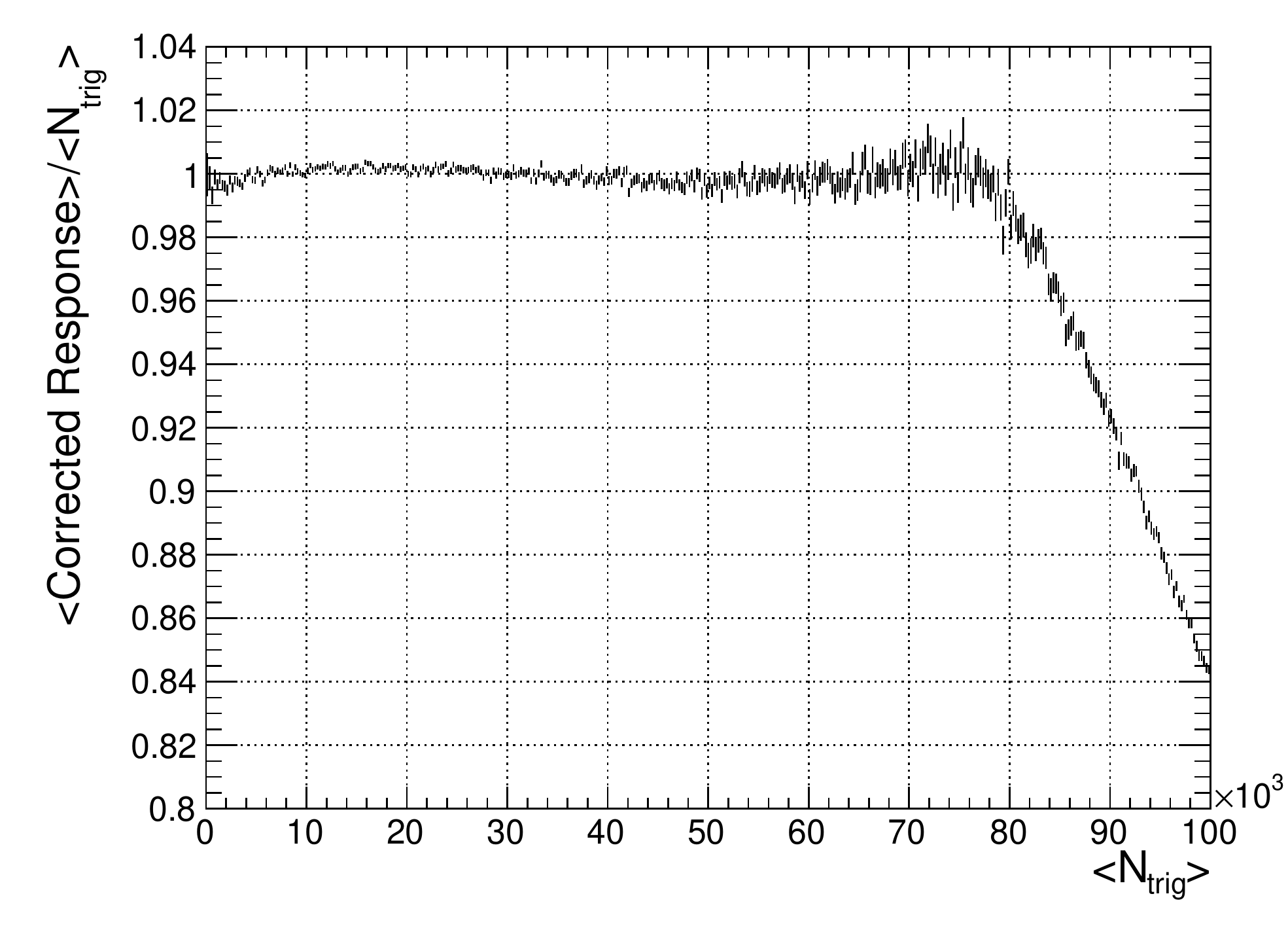}
    \includegraphics[totalheight=0.2\textheight]{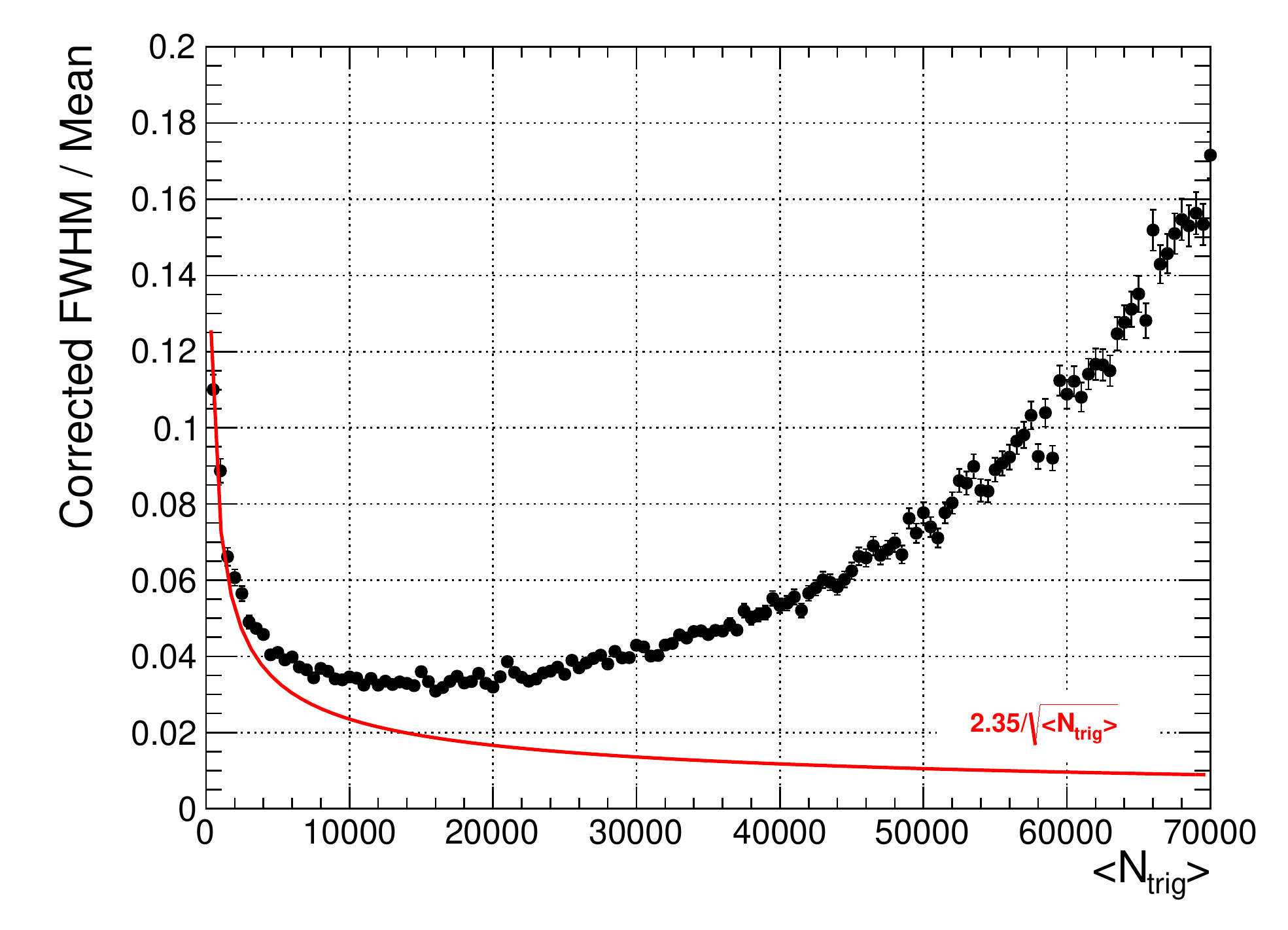}
    \caption{\textbf{Left:} After corrections the mean response has the correct scale within 0.5\% up to $\sim 70,000$ triggering photons. \textbf{Right:} The pulse height resolution after corrections versus the number of triggering photons.
The resolution includes the contribution from the statistical fluctuations of
the number of triggering photons shown by the red line.
It is assumed that the number of incoming photons is described by the Poisson
distribution.        }
\label{fig:correction}
\end{figure}

The response of the detector is basically given by the number of fired
microcells multiplied by the charge released by a single microcell. Small random fluctuations of the charge released by individual microcells (typically, about 10\%) smear and slightly widen the response distribution. Small coherent charge (gain) fluctuations e.g. due to bias voltage fluctuations, can have a more significant effect. The detector characteristics used in our simulations (Table~\ref{sim_params}) are assumptions based on the
performance of the current SensL devices and the expected PDE of
the next-generation, blue sensitive, devices. Due to the relatively long microcell recovery time we
consider the incident light to be instantaneous (the scintillation decay time in LaBr$_3$ is $\sim20$\,ns and the dispersion of the light transportation time in the considered detector configuration is $\sim5$\,ns). For the same reason the after-pulse effect is neglected
in our model. Results of the response simulations are shown in Figures~\ref{fig:response} and ~\ref{fig:correction}.

\begin{table}
  \begin{tabular}{|c|c|c|c|c|c|} \hline
    Number of   & PDE               & Dark  & Microcell &
    Cross-talk  &  Gain stability \\
    microcells  & at 340-400\,nm  & count rate & recovery time&
    probability & (RMS) \\
    \hline
    18980 & 20\% & 30 MHz & 100 ns & 20\% & 0.5\%  \\
    \hline
  \end{tabular}
       \caption{SiPM parameters assumed in the simulations. The performance characteristics are given at the nominal bias voltage (2V above the breakdown voltage) and a temperature of 20$^\circ$C.}
   \label{sim_params}
\end{table}

\section{Calorimeter module design}

\begin{figure}[htb]
  \centering
  \begin{minipage}[c]{0.58\textwidth}
   \centering
   \includegraphics[trim=1cm 4cm 1cm 4cm, clip=true, totalheight=0.4\textheight]{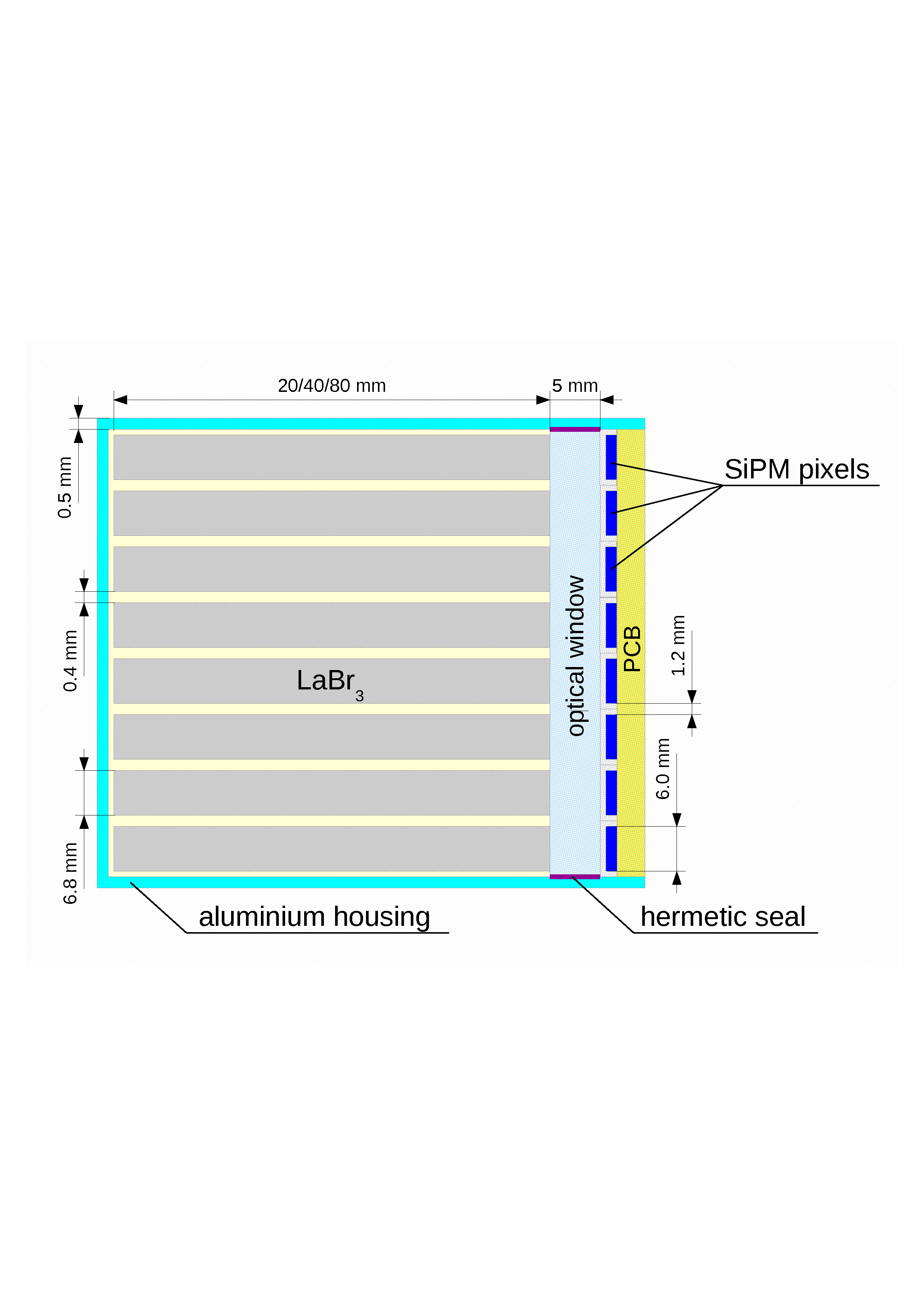}
  \end{minipage}
  \begin{minipage}[c]{0.38\textwidth}
   \centering
      \begin{footnotesize}
          \begin{tabular}{|l|}
  \hline
  Polished LaBr$_3$ crystals with n$_{\rm refraction}$= 2.3 \\
 Optical grease with n$_{\rm refraction}$= 1.46 \\ between
crystals/window/sensors\\
 No light absorption in crystals \\
 Light scattering length of 90-150 mm \\
depending on wavelength \\
Diffuse reflector around crystals with 98\% \\
reflectivity (with air gap) \\
 Hermetic seal with 70\% diffuse reflectivity \\
  \hline
        \end{tabular}
     \end{footnotesize}
  \end{minipage}
\caption{Geant4 model of the calorimeter module (left). The module includes a hermetically sealed $8\times8$ crystal array and a matching array of SiPM pixels. The optical model properties are indicated on the right.}
  \label{model}
\end{figure}

LaBr$_3$ is a unique scintillator with very high light yield ($\sim 60,000$~photons/MeV) and short decay time ($\sim 20$\,ns). An energy resolution better than 3\% at 662\,keV has been measured with photomultiplier tubes.
However, LaBr$_3$ is very hygroscopic and requires an air-tight enclosure, which is a major complication in the design of a detector. A possible design of the GRIPS calorimeter module is shown in Figure~\ref{model}. The 5\,mm thickness of the optical window is motivated by the sealing requirement.

\begin{figure}
    \includegraphics[trim=1cm 1cm 1cm 1cm, clip=true, totalheight=0.3\textheight, angle=-90]{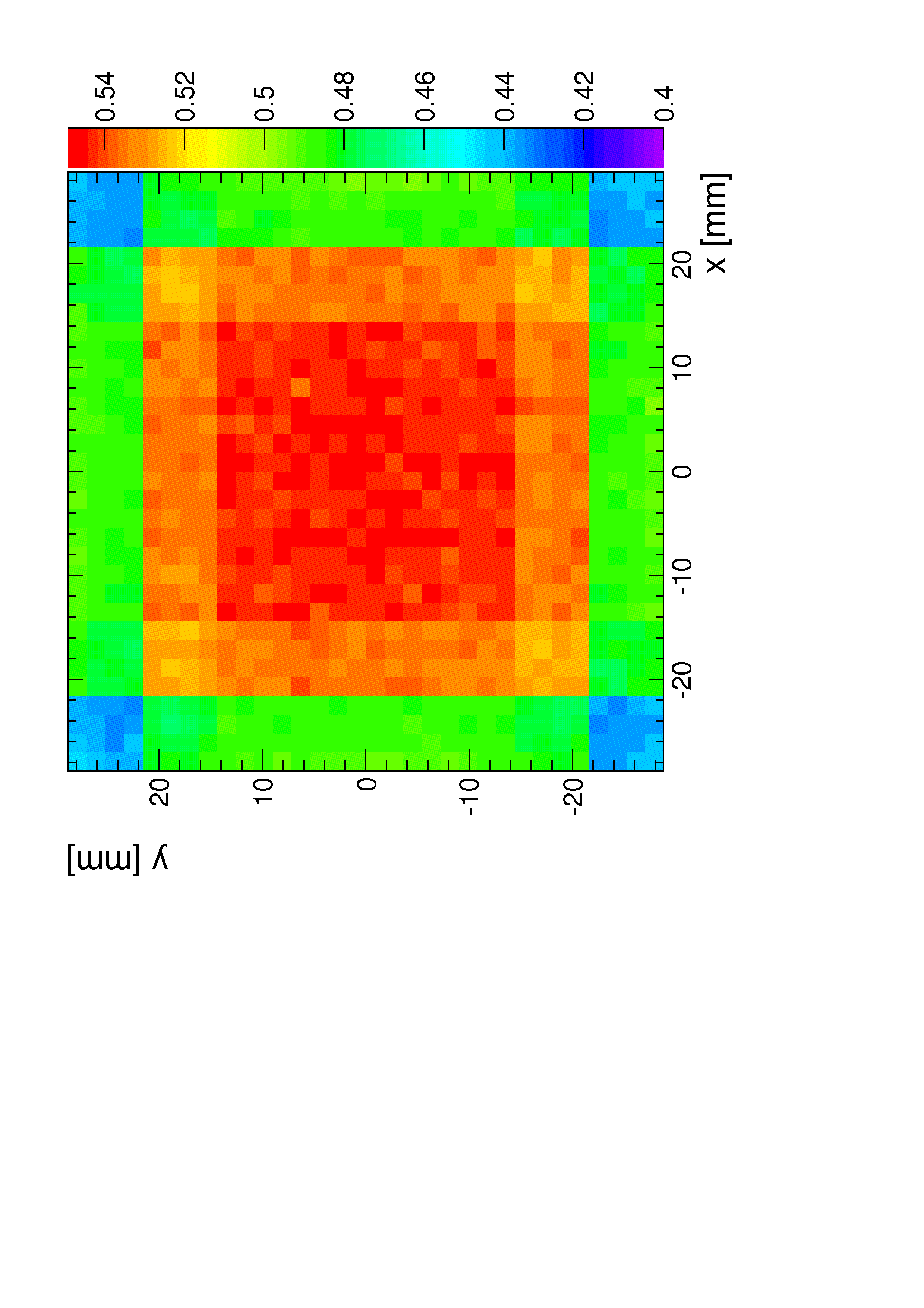}
    \includegraphics[trim=1cm 1cm 1cm 1cm, clip=true, totalheight=0.3\textheight, angle=-90]{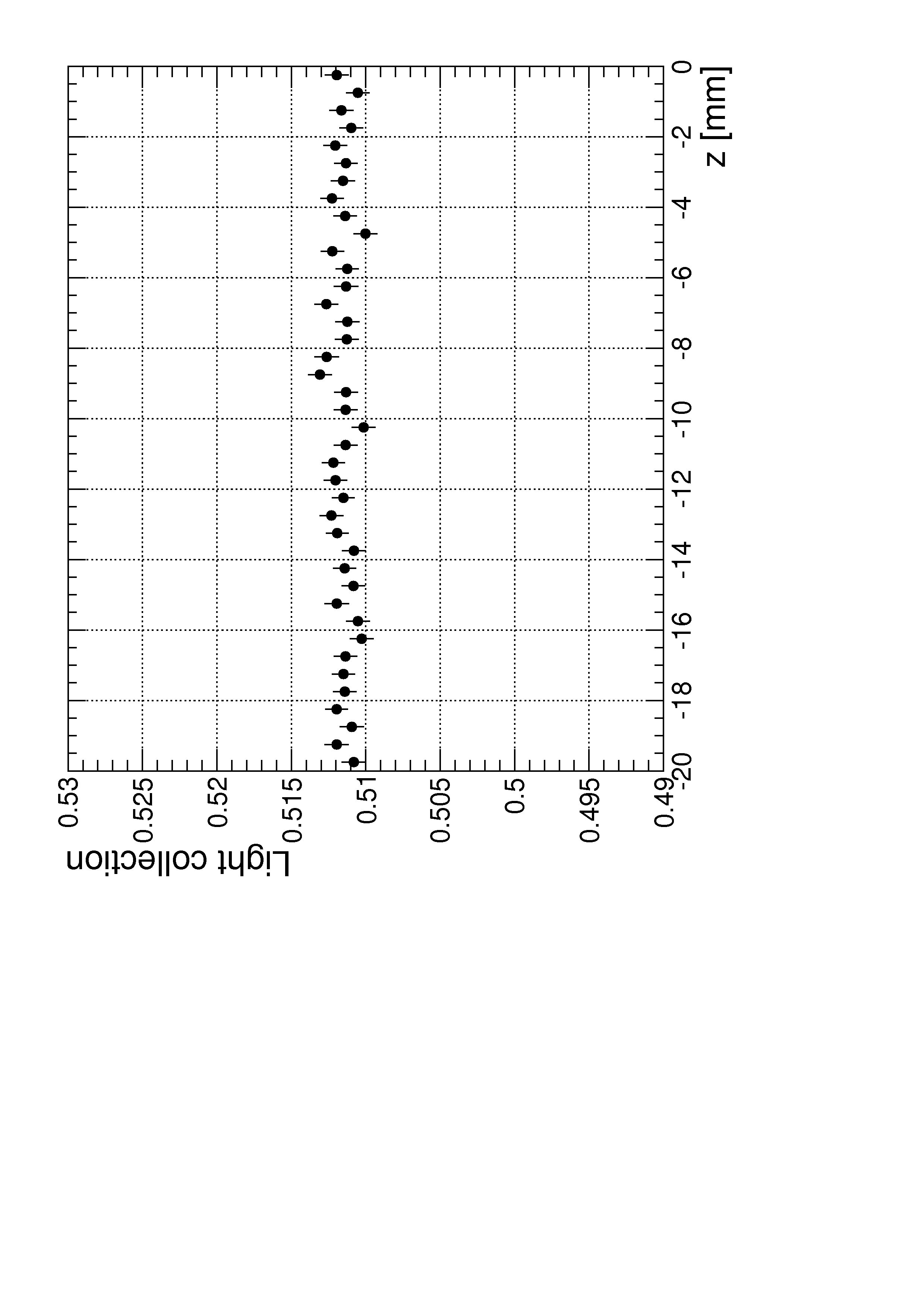}
    \caption{The light collection efficiency of the module versus the transverse (\textbf{left}) and longitudinal (\textbf{right}) coordinates of the light emission point. }
\label{fig:light_collection}
\end{figure}

\begin{figure}
    \includegraphics[trim=1cm 1cm 1cm 1cm, clip=true, totalheight=0.3\textheight, angle=-90]{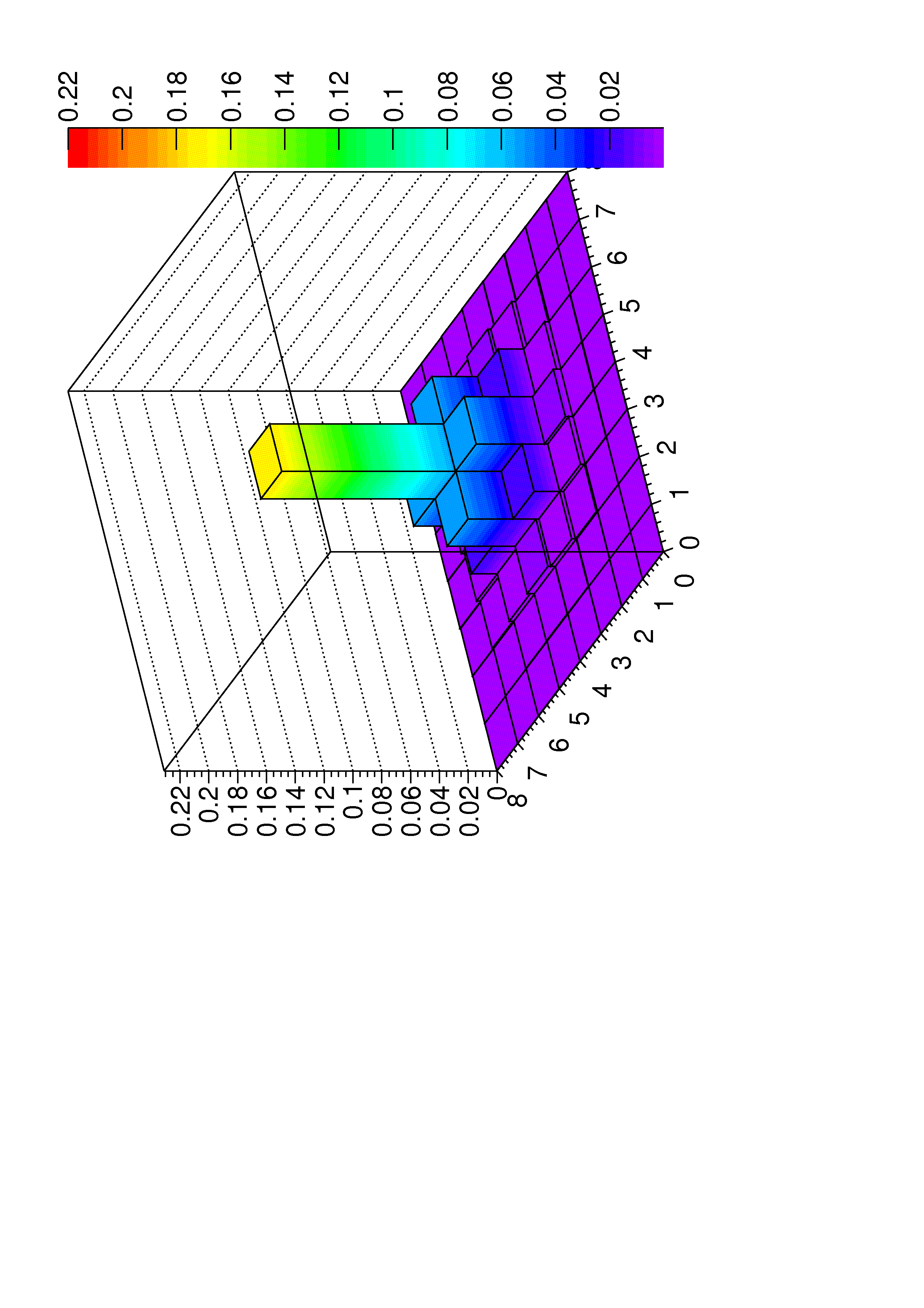}
    \includegraphics[trim=1cm 1cm 1cm 1cm, clip=true, totalheight=0.3\textheight, angle=-90]{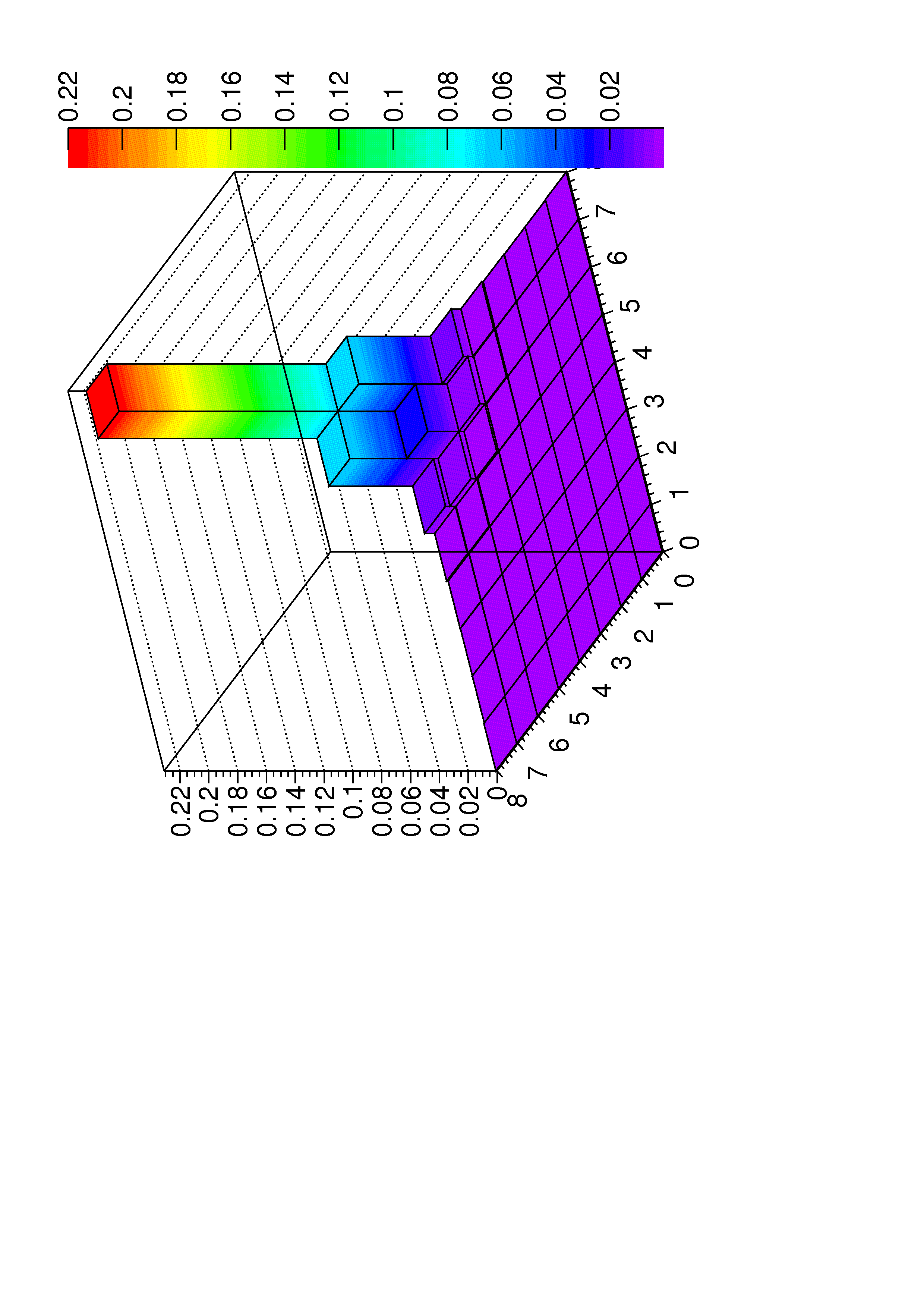}
    \caption{Light distribution among the 64 detector pixels for the light produced in a central crystal (left) and in a corner crystal (right). The distributions are normalised to give the generated light fractions per pixel.}
\label{fig:light_distribution}
\end{figure}

This design has been modelled with the Geant4 toolkit~\cite{geant4} to simulate the transportation of light through the module. In the model we use the optical properties of LaBr$_3$ recently reported in~\cite{labr3optical}. Although the light attenuation length was measured in that work, the authors suggest that the re-emission probability of the absorbed photons is close to unity. It is not known if any significant losses of the scintillation light happen in the bulk of LaBr$_3$ crystals. Therefore, we do not consider light absorption by LaBr$_3$ in the simulations but the finite reflectivities of the crystal wrapping and the optical window seal are taken into account. Results of the optical simulations are shown in Figures~\ref{fig:light_collection} and~\ref{fig:light_distribution}.

The thick optical window results in considerable optical cross-talk between detector pixels. When light is produced in a central crystal, only 17\% of the light is collected by the matching detector pixel (32\% of the light collected by all pixels). $3\times 3$ pixels get 47\% (86\%) of the light. In the case of a corner crystal the single pixel light fraction is increased to 21\% (50\%) due to light reflections from the window edges. In spite of the cross-talk the position of the emitting crystal remains easily identifiable even for small absorbed energies.  Assuming PDE=20\%, a single SiPM would get 2400 triggering photons per 1\,MeV deposit in the matching crystal. From Figure~\ref{fig:correction} we conclude that the modelled detector can measure the energy absorbed in a single crystal up to about 30\,MeV (72,000 triggering photons).

\section {Future work}
The next steps in this project include:
\begin{itemize}
\item Laboratory tests of the SiPMs supplied by SensL together with LaBr$_3$ crystals.
\item Implementation of our module design and SiPM response model in the existing GRIPS simulation code which is based on Geant4 and MEGAlib~\cite{megalib}. Given the energy range covered by GRIPS and the huge number of photons produced it is not feasible to track individual photons in GRIPS simulations. Instead we intend to use parametrised light distributions similar to those shown in Figure~\ref{fig:light_distribution}.
\item Explore other possible module configurations such as a monolithic scintillator read out by an array of SiPM pixels.
\end{itemize}

\section*{Acknowledgements}

We would like to thank Carl Jackson, Kevin O'Neill and Paul Sheridan from SensL for collaboration on this project.
This work is being supported under ESA's Strategic Initiative AO/1-6418/10/NL/Cbi.


\begin{thebibliography}{99}
\bibitem{greiner2009} J. Greiner et al.: Gamma-ray burst investigation via polarimetry and spectroscopy (GRIPS), Exp. Astron. 23, 91 (2009).

\bibitem{greiner}
J. Greiner et al.: GRIPS - Gamma-Ray Imaging, Polarimetry and Spectroscopy. arXiv:1105.1265 (2011).

\bibitem{garutti2011}
E. Garutti: Silicon photomultipliers for high energy physics detectors, Journal of Instrumentation, Vol. 6 (2011).

\bibitem{labr3optical}
H. T. van Dam et al.: Optical Absorption Length, Scattering Length, and
Refractive Index of LaBr$_3$:Ce3+, IEEE Trans. Nucl. Sc. 59, 656 (2012).

	


\bibitem{geant4}
S. Agostinelli et al.: Geant4 - a simulation toolkit. Nucl. Instr. Meth. A506, 250 (2003).

\bibitem{megalib}
A. Zoglauer, R. Andritschke, F. Schopper: MEGAlib - the medium energy gamma-ray astronomy library, New Astron. Rev. 50(7-8), 629 (2006).




\end{thebibliography}
\end{document}